\documentclass{article}

\usepackage{PRIMEarxiv}

\usepackage[utf8]{inputenc} 
\usepackage[T1]{fontenc}    
\usepackage[colorlinks = true,
        urlcolor  = blue]{hyperref}
\usepackage{url}            
\usepackage{booktabs}       
\usepackage{amsfonts}       
\usepackage{nicefrac}       
\usepackage{microtype}      
\usepackage{lipsum}
\usepackage{fancyhdr}       
\usepackage{graphicx}       
\graphicspath{{media/}}     

\usepackage{amsmath}
\usepackage{booktabs}
\usepackage{hyperref}
\usepackage{subcaption}
\usepackage{enumitem}
\usepackage{listings}
\usepackage{xurl}

\title{Digital Collections Explorer: An Open-Source, Multimodal Viewer for Searching Digital Collections}

\author{
Ying-Hsiang Huang \\
Information School \\
University of Washington \\
Seattle, WA, USA
\And 
Benjamin Charles Germain Lee \\
Information School \\
University of Washington \\
Seattle, WA, USA \\
\texttt{bcgl@uw.edu}
}

\begin{document}
\maketitle

\begin{abstract}
We present Digital Collections Explorer, a web-based, open-source exploratory search platform that leverages CLIP (Contrastive Language-Image Pre-training) for enhanced visual discovery of digital collections. Our Digital Collections Explorer can be installed locally and configured to run on a visual collection of interest on disk in just a few steps. Building upon recent advances in multimodal search techniques, our interface enables natural language queries and reverse image searches over digital collections with visual features. This paper describes the system's architecture, implementation, and application to various cultural heritage collections, demonstrating its potential for democratizing access to digital archives, especially those with impoverished metadata. We present case studies with maps, photographs, and PDFs extracted from web archives in order to demonstrate the flexibility of the Digital Collections Explorer, as well as its ease of use. We demonstrate that the Digital Collections Explorer scales to hundreds of thousands of images on a MacBook Pro with an M4 chip. Lastly, we host a public demo of Digital Collections Explorer.
\end{abstract}


\section{Introduction}
Despite the significant advances in providing access to both digitized and born-digital collections over the past three decades, digital collections -- particularly those with visual features -- face significant challenges surrounding discoverability. While manually-curated metadata for photographs, maps, and other visual culture are incredibly valuable when searching a collection, this approach simply does not scale to millions of items. The digitized \textit{Chronicling America} newspaper collection now has over 20 million individual pages digitized, and born-digital collections are even larger, with petabytes of data comprising billions of items. As a result, collections often lack basic descriptive metadata -- and without basic metadata facets, it is fundamentally difficult to search collections. 

Researchers in the computational humanities and cultural heritage have long been interested in automated approaches to metadata augmentation, as evidenced by the long history of optical character recognition (OCR) for the text transcription of digitized documents \cite{cordell_2020}. The advent of multimodal models such as CLIP \cite{radford_2021} that capture visual and textual information jointly have shown great promise for addressing this challenge for collections ranging from maps \cite{mahowald_lee} to newspapers \cite{smits_JOHD}. While this research has demonstrated the ability to search over collections with little to no associated metadata, this research must still be translated into practice. Stewards of these collections are in need of democratized solutions for making their collections discoverable using these approaches -- in particular, ones that are accessible to non-experts with access to only standard, staff-issued hardware (e.g., a MacBook).

In this paper, we introduce our Digital Collections Explorer, an open-source framework that can be run locally on a laptop with only a few steps in order to spin up a multimodal search interface for a digital collection with hundreds of thousands of items. With the Digital Collections Explorer, end-users can interactively search large-scale collections using multiple input modalities, including both natural language inputs (e.g., ``redacted documents'') and visual inputs (i.e., reverse image search). Our inspiration for and implementation of the Digital Collections Explorer is based on extended collaborations with stewards of collections who have articulated precisely these needs.

\begin{figure}[t]
    \centering
    \includegraphics[width=\linewidth]{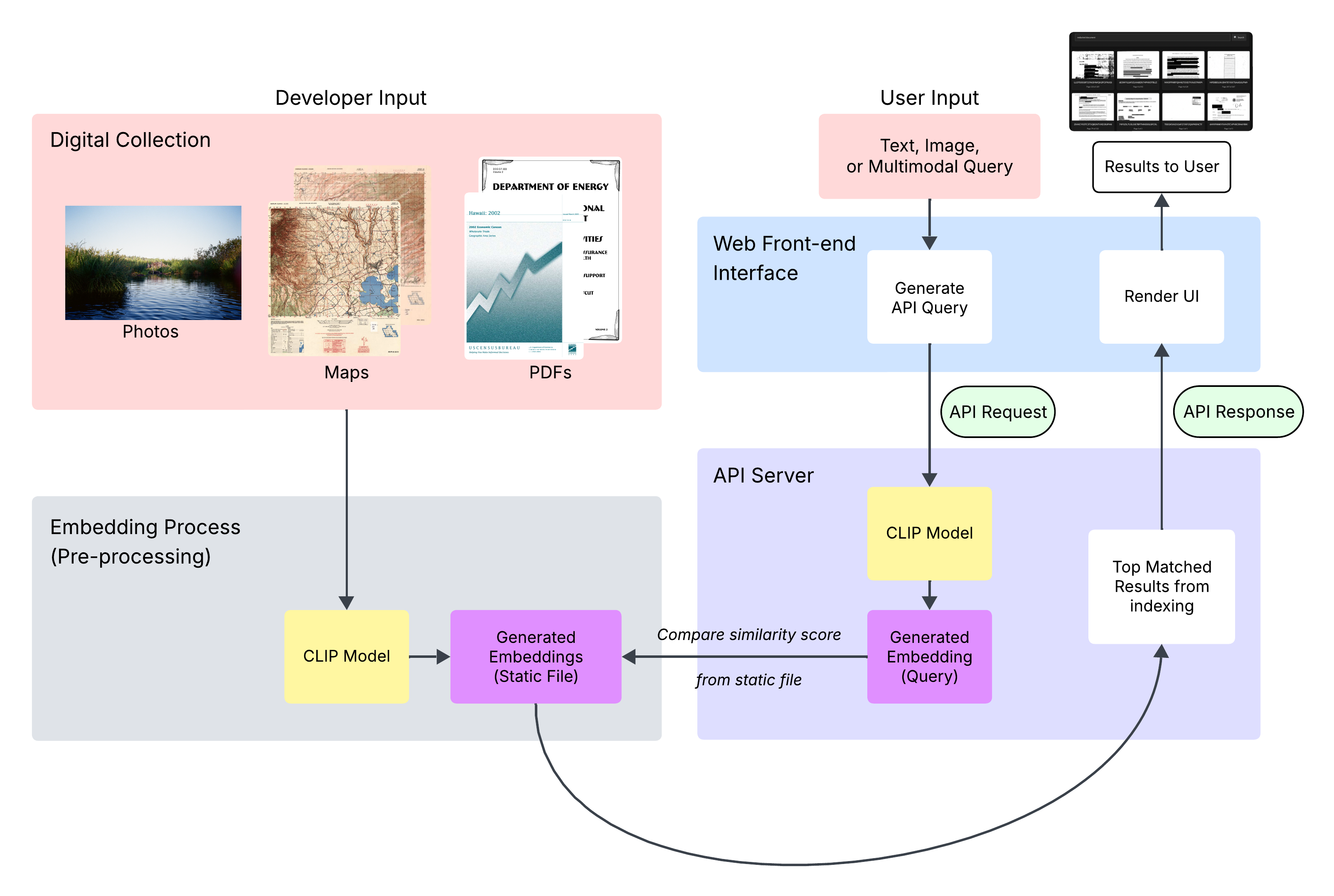}
    \caption{
    An overview of the Digital Collections Explorer, showing the central components: 1) the developer input, 2) the embedding process, 3) user input, 4) the web front-end interface, and 5) the API server. 
    }
    \label{fig:overview}
\end{figure}

The Digital Collections Explorer is designed to be easy to use for non-experts and extensible to a wide range of collections with visual features, from visual culture to documents with visual layouts and other semantic features encoded visually. In Figure~\ref{fig:overview}, we show an overview of how the Digital Collections Explorer works. To spin up the Digital Collections Explorer, the developer inputs a digital collection (red, top-left). This initiates the embedding process (gray, bottom-left), which generates CLIP embeddings for all of the items in the collection. Based on our case studies presented in this paper, this step can scale to hundreds of thousands of items on a personal laptop. Once the embedding pre-processing is finished, the developer can spin up the viewing interface (blue, top-right), which:
\begin{enumerate}
    \item takes a user's searches as input (red, top-right)
    \item communicates with the API server (violet, bottom-left) to embed the search query and identify the top results using the CLIP embeddings
    \item renders the front-end interface and displays the results to the user (blue, top-right).
\end{enumerate} 
Our Digital Collections Explorer is designed to be run end-to-end locally, meaning that the embedding pipeline utilizes a locally-installed model (without transferring a digital collection to any external API), and the viewer can be spun up on a local machine as well, without being made publicly visible. In this regard, the Digital Collections Explorer can be applied even to digital collections with sensitivities surrounding privacy and access. Users interact with the system through a React-based front-end, which supports natural language queries, reverse image search, and multimodal inputs. The back-end, implemented in FastAPI, handles search requests by comparing query embeddings with precomputed image embeddings. 

To aid those interested in using our software, we provide a tutorial for running the Digital Collections Explorer with the goal of facilitating use by researchers and practitioners in the computational humanities, as well as in galleries, libraries, archives, and museums. We demonstrate the extensibility of our Digital Collections Explorer with four different collections: two photojournalism collections provided to us by collaborators due to the collections' lack of descriptive metadata (and thus persistent difficulties searching them); a collection of 562,842 images of maps held by the Library of Congress; and a collection of a thousand born-digital PDFs produced by the federal government. In doing so, we demonstrate how the Digital Collections Explorer can facilitate searching even in the limit of no metadata. Lastly, to demonstrate the functionality of the Digital Collections Explorer, we host a public demo at \url{https://www.digital-collections-explorer.com} for searching these 500,000+ images of maps from the Library of Congress.

\subsection{Contributions}

This paper presents several contributions:

\begin{enumerate}
    \item We introduce our Digital Collections Explorer, an open-source cultural heritage viewer for visual culture exploration. The system provides institutions with a robust foundation for digital collection management and discovery, while addressing key challenges in user interaction. Significantly, our Digital Collections Explorer can be spun up locally, meaning that both the machine learning embedding pipeline and the viewer can be spun up without making any data visible to the public or to any machine learning APIs.
    
    \item Our Digital Collections Explorer implements a metadata-agnostic approach, enabling semantic search and exploration capabilities even for collections lacking traditional metadata structures. By leveraging CLIP embeddings, this approach significantly expands the accessibility of previously hard-to-search archival materials.
 
    \item Our research contributes to the open-source community through a comprehensive implementation, available via a public repository, as well as our tutorial for applying our Digital Collections Explorer to other collections of interest. The codebase is publicly available at \url{https://doi.org/10.5281/zenodo.15744570} and is available with a CC-BY-4.0 license. 

    \item We demonstrate the Digital Collections Explorer's adaptability across diverse collection types, including photographs, maps, and born-digital documents.
  
    \item We host a public demo of Digital Collections Explorer on an example collection of 562,842 digitized map images from the Library of Congress at: \url{https://www.digital-collections-explorer.com}.
\end{enumerate}

\section{Related Work}

In this section, we contextualize our work in relation to existing projects and literature surrounding the collections as data, multimodal cultural heritage, and open-source viewers for digital cultural heritage.

\subsection{Collections as Data and Responsible AI}
We build on extensive work over the past decade to develop ``Collections as Data'' approaches \cite{collections_as_data_2017, collections_as_data_2019}. ``Collections as Data'' principles emphasize ``computational use of digitized and
born digital collections,'' ``lower[ing] barriers to use,'' ``shared documentation help[ing] others find a path to doing the work,'' and ``valu[ing] interoperability'' \cite{collections_as_data_2019}, all of which are principles that we bring with our Digital Collections Explorer. 

We also draw from the related area of work surrounding responsible uses of AI for galleries, libraries, archives, and museums (GLAMs) \cite{padilla_responsible_2020, collections_as_ml_data, labs_fw}. In particular, we have drawn from this literature during our development process and have chosen to emphasize the development of tooling that uses AI in order to improve access and democratize its application, while also ensuring that privacy and stewardship are emphasized through the adoption of open models and local interfaces.

\subsection{Multimodal Cultural Heritage}

Our Digital Collections Explorer builds upon a rich body of research in multimodal search and digital cultural heritage. Recent advancements in multimodal machine learning have yielded the development of open models such as CLIP \cite{radford_2021} and more recently, LlaVa \cite{llava_2024} and Molmo \cite{molmo}. These models have enabled semantic alignment between text and image embeddings, facilitating a wealth of searches across language and vision. Prior work has explored the application of these models to cultural heritage collections \cite{mahowald_lee, smits_JOHD, smits_wevers, smits_kestemont, wevers_photos} and has demonstrated promising possibilities for improving the discoverability of large-scale visual collections, especially those with little descriptive metadata. However, challenges remain in democratizing these approaches and integrating them into user-friendly systems for viewing. Our Digital Collections Explorer addresses this challenge by prioritizing extensiblity for non-experts. In the tradition of open machine learning models that can be run locally -- without sharing information with proprietary AI companies -- our Digital Collections Explorer is designed to use open multimodal models, ensuring that digital collections can be stewarded properly.

\subsection{Open-source Viewers for Digital Collections}

Researchers and practitioners in digital cultural heritage have long contributed to the creation of open-source image viewing software for digital collections, enabling non-experts to spin up interfaces for viewing. Viewers such as CollectionBuilder \cite{collectionbuilder} and Omeka \cite{omeka} provide faceted viewing options and have been heavily utilized within the digital humanities, computational humanities, and library communities. Viewers such as PixPlot \cite{pixplot}, CollectionScope \cite{collectionscope}, and artexplorer.ai \cite{artexplorer} support visual and multimodal semantic search in a more exploratory fashion via cluster-based search. Other innovative viewers include the Vikus Viewer \cite{vikus}. Our Digital Collections Explorer builds on this movement in order to provide new modes of open-source viewing. Our solution, which is released with a CC-BY-4.0 license, is designed with both extensibility and scale in mind, providing semantic viewing capabilities over hundreds of thousands of items seamlessly.  
\begin{figure*}
    \centering
    \includegraphics[width=\linewidth]{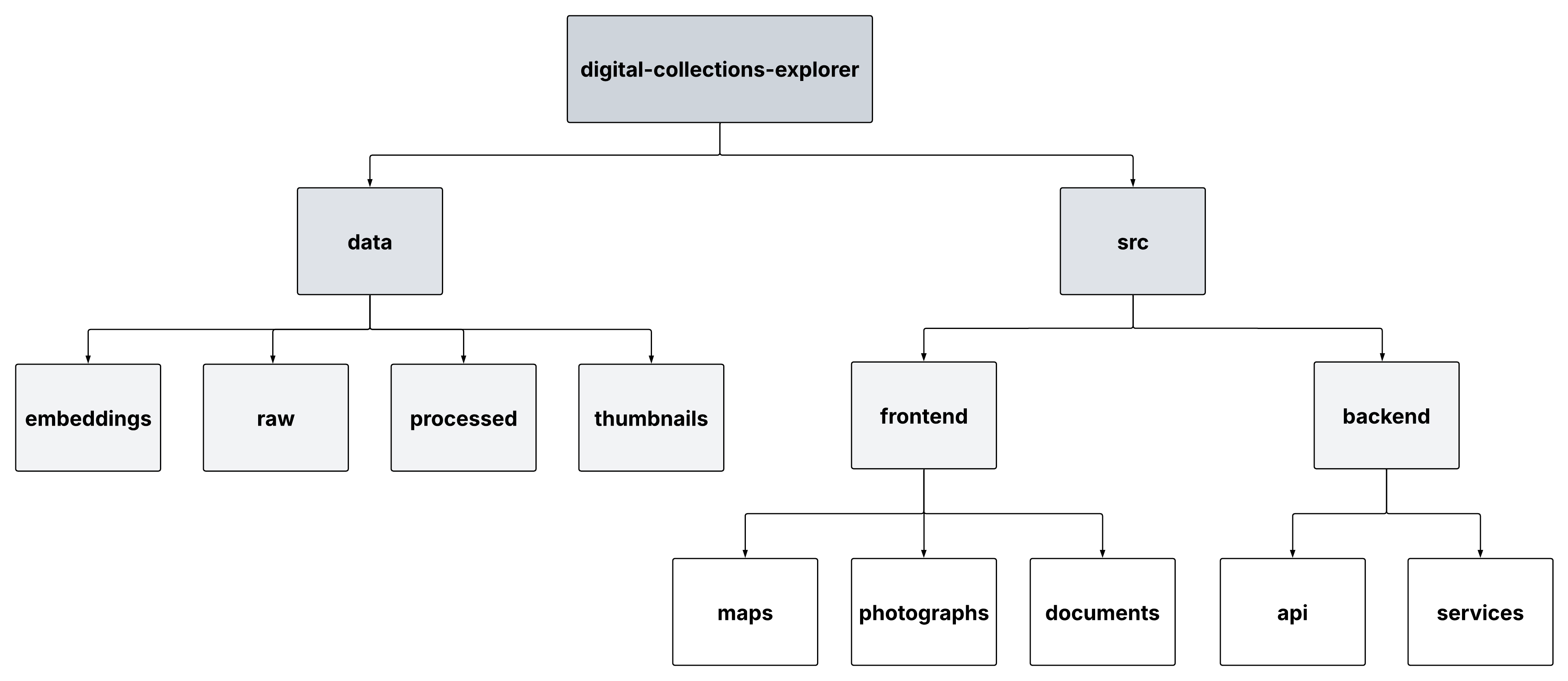}
    \caption{The directory layout of the Digital Collections Explorer codebase, showing the high-level organization into \texttt{data}, \texttt{src/frontend}, and \texttt{src/backend}, which correspond to distinct functional layers of the system.}
    \label{fig:codebase_structure_overview}
\end{figure*}

\section{Digital Collections Explorer: An Overview}

In this section, we include an overview of our system architecture, including the embedding generation pipeline, front-end, and back-end components, as well as a tutorial describing how to spin up a local instance of the Digital Collections Explorer, along with a public demo.

\begin{figure*}[t]
    \centering
    \includegraphics[width=0.6\linewidth]{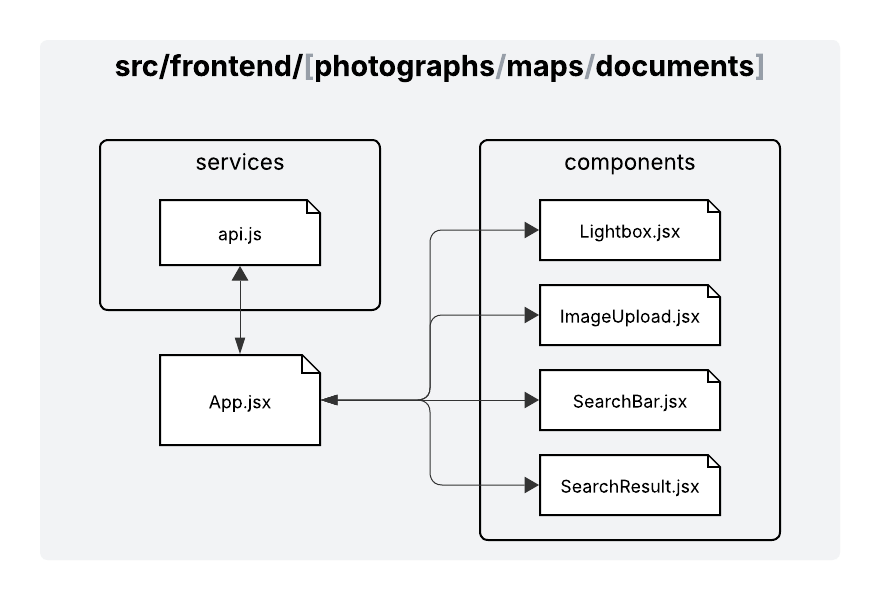}
    \caption{The component-based front-end architecture of the Digital Collections Explorer. The parent (\texttt{App.jsx}) component acts as a stateful controller, mediating between the reusable UI and an abstracted layer (\texttt{api.js}).}
    \label{fig:codebase_fe_structure_overview}
\end{figure*}

\subsection{System Architecture}

The Digital Collections Explorer is designed as a modular system, ensuring maintainability, scalability, and ease of reuse. The overall structure of the codebase is illustrated in Figure~\ref{fig:codebase_structure_overview}, consisting of three central branches: \texttt{data}, \texttt{src/frontend}, and \texttt{src/backend}.

\subsubsection{Embedding Generation Pipeline}

The system uses a local implementation of CLIP to generate embeddings of visual collections. This ensures privacy because no data is sent to external servers, making the system suitable for sensitive collections.
Likewise, it ensures efficiency, as embeddings are generated locally, reducing dependency on external APIs and ensuring consistent performance.
By default, the system loads the publicly available pre-trained model:\footnote{
    The model card for \texttt{clip-vit-base-patch32} can be found at: \url{https://huggingface.co/openai/clip-vit-base-patch32}
}
\begin{center}
    \texttt{openai/clip-vit-base-patch32} 
\end{center}

The generation pipeline is managed through a set of clearly defined directories within the \texttt{data} folder, as illustrated in Figure~\ref{fig:codebase_structure_overview}, which is systematically organized as follows:
    \begin{itemize}
        \item \textbf{\texttt{raw}}: This directory serves as the initial input location for the user's original collection files in their native format (e.g., JPGs, PNGs, TIFFs, or PDFs).
        \item \textbf{\texttt{processed}}: Before embedding, certain files require pre-processing. For instance, PDFs are converted into a series of images during pre-processing, and these intermediate files are stored here.
        \item \textbf{\texttt{thumbnails}}: To ensure a smooth user experience in the gallery view, the system automatically generates low-resolution thumbnails for each item, which are stored in this directory for rapid loading.
        \item \textbf{\texttt{embeddings}}: The final output of the pipeline, the computed tensor embeddings, are saved as \texttt{.pt} files in this directory. Instead of generating embeddings with Digital Collections Explorer, users may skip the pipeline and place their pre-computed embeddings here for the system to use directly.\footnote{To use pre-computed embeddings, three conditions must be met:
                \begin{enumerate}[label=\arabic*), topsep=0pt, itemsep=2pt, parsep=0pt, partopsep=0pt, leftmargin=4em]
                    \item An \texttt{embeddings.pt} file containing a single PyTorch tensor of shape \texttt{[N, D]}, where \texttt{N} is the total number of items and \texttt{D} is the embedding dimension.
                    \item An \texttt{item\_ids.pt} file containing a Python list of \texttt{N} unique string identifiers.
                    \item The dimensionality of the custom embeddings must precisely match the output dimension of the model specified in the \texttt{config.json}.
                \end{enumerate}
            }
    \end{itemize}
The model choice is fully configurable: users can swap in any Hugging Face transformers-compatible model by editing a single line in the project’s \texttt{config.json}.
As described in the tutorial later in this section, once a digital collection is placed in the \texttt{raw} directory, the embedding pipeline can be run with a single command.

\subsubsection{Front-end}

\begin{figure*}[t!]
    \centering
    \begin{subfigure}[b]{0.48\textwidth}
        \centering
            \includegraphics[height=1.7in]{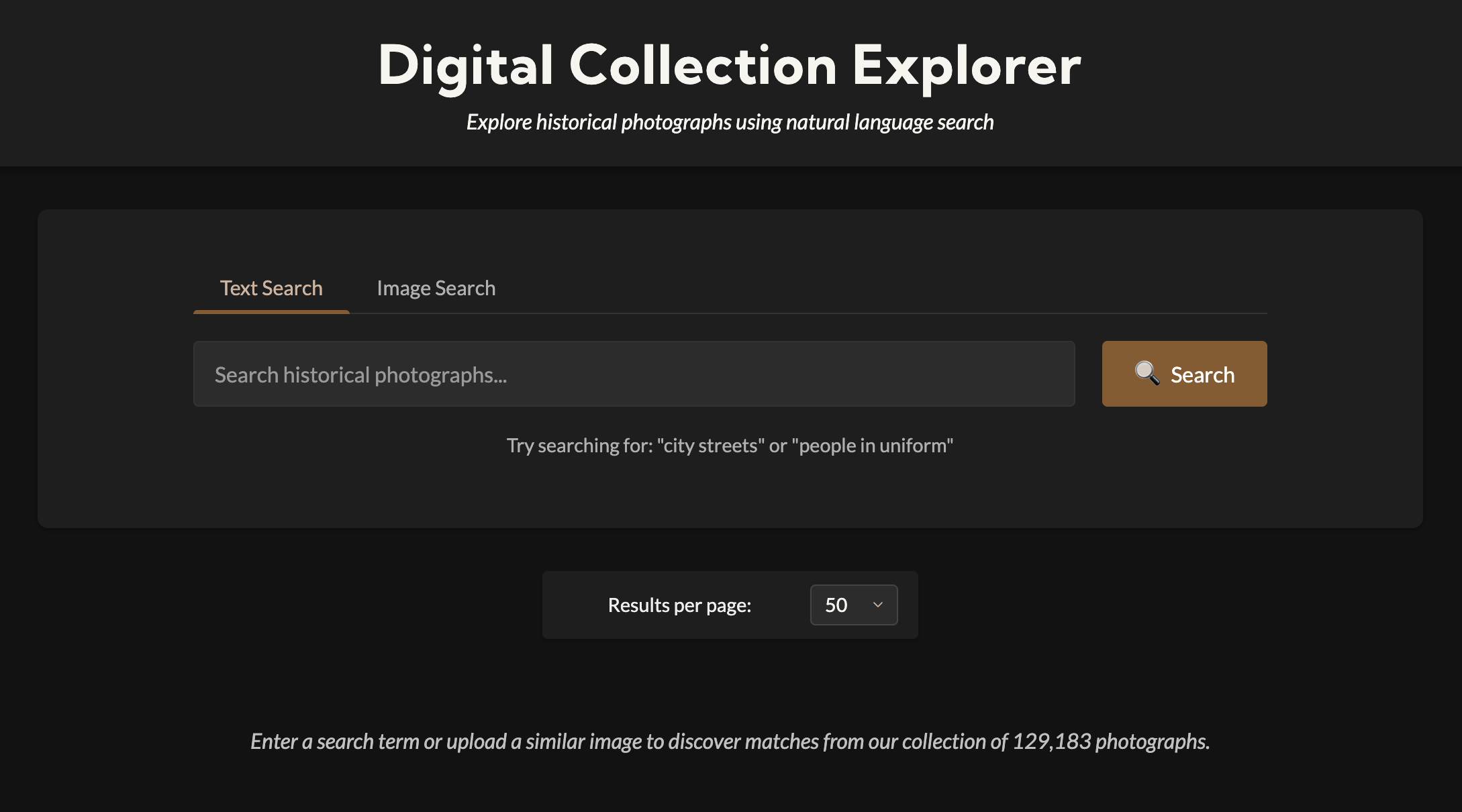}
        \caption{Text search.}\label{fig:text}
    \end{subfigure}
    ~ 
    \begin{subfigure}[b]{0.48\textwidth}
        \centering
            \includegraphics[height=1.7in]{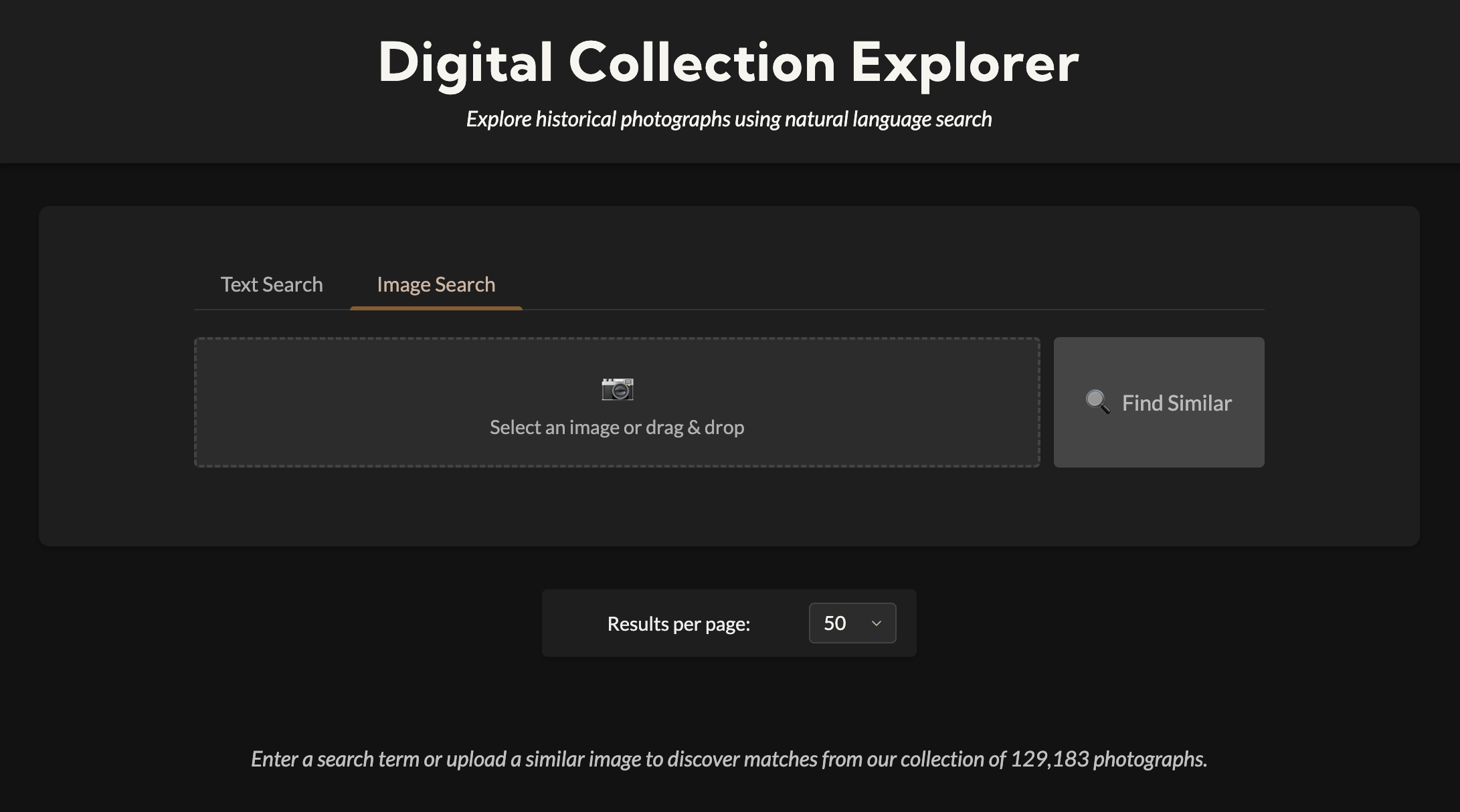}
        \caption{Image search.}\label{fig:image}
    \end{subfigure}
    \caption{Examples of the landing page for the photographs collection interface, which presents an end-user with two options for searching: text search via natural language  (Figure \ref{fig:text}) and image search (Figure \ref{fig:image}).}\label{fig:interface_demo_01}
\end{figure*}

\begin{figure}[t]
    \centering
    \includegraphics[width=\linewidth]{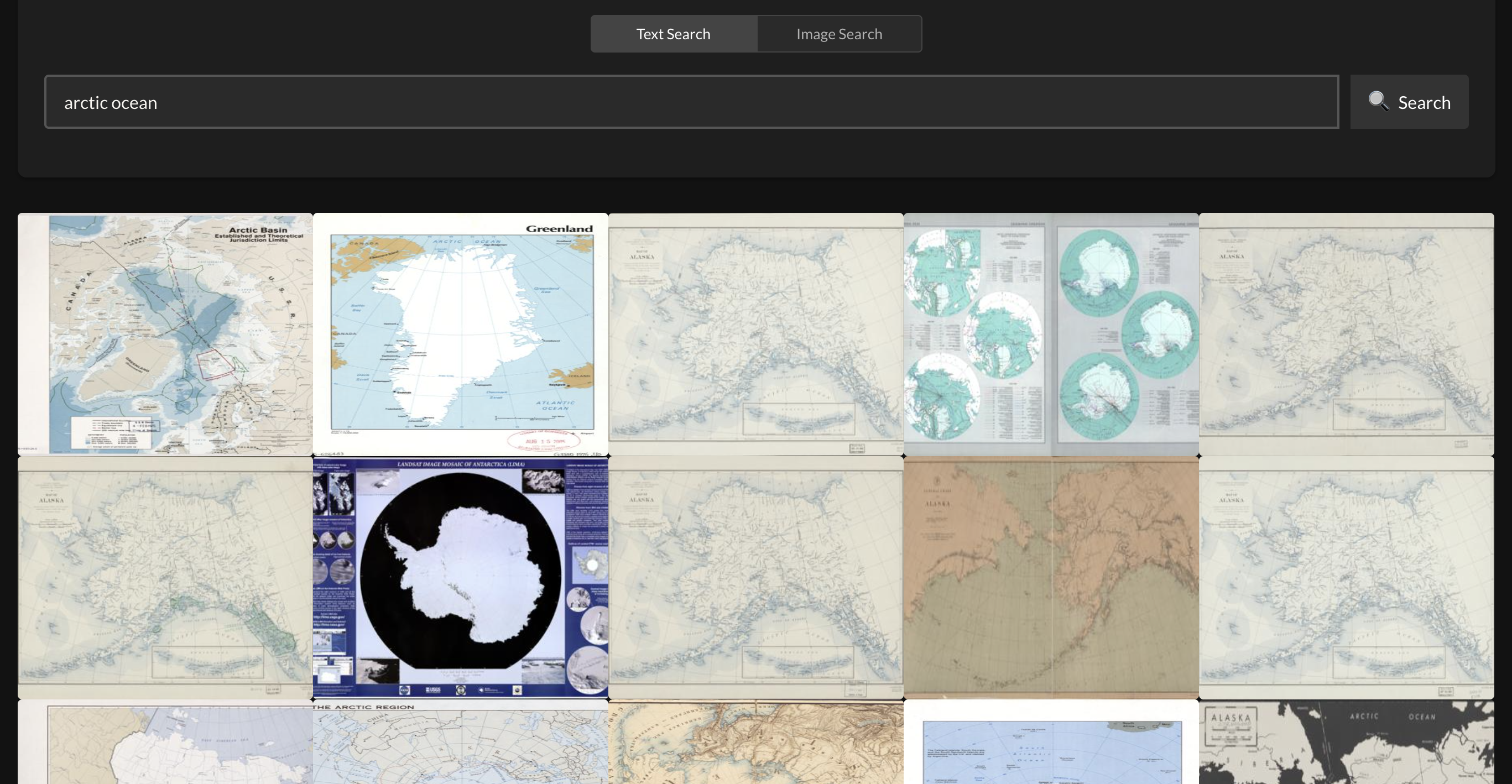}
    \caption{An example of the historical maps gallery view rendered by the \texttt{SearchResult.jsx} component in response to a user query ``arctic ocean''. The component's responsibility is to render a grid with thumbnails of the maps.}
    \label{fig:interface_demo_02}
\end{figure}

\begin{figure}[t]
    \centering
    \includegraphics[width=\linewidth]{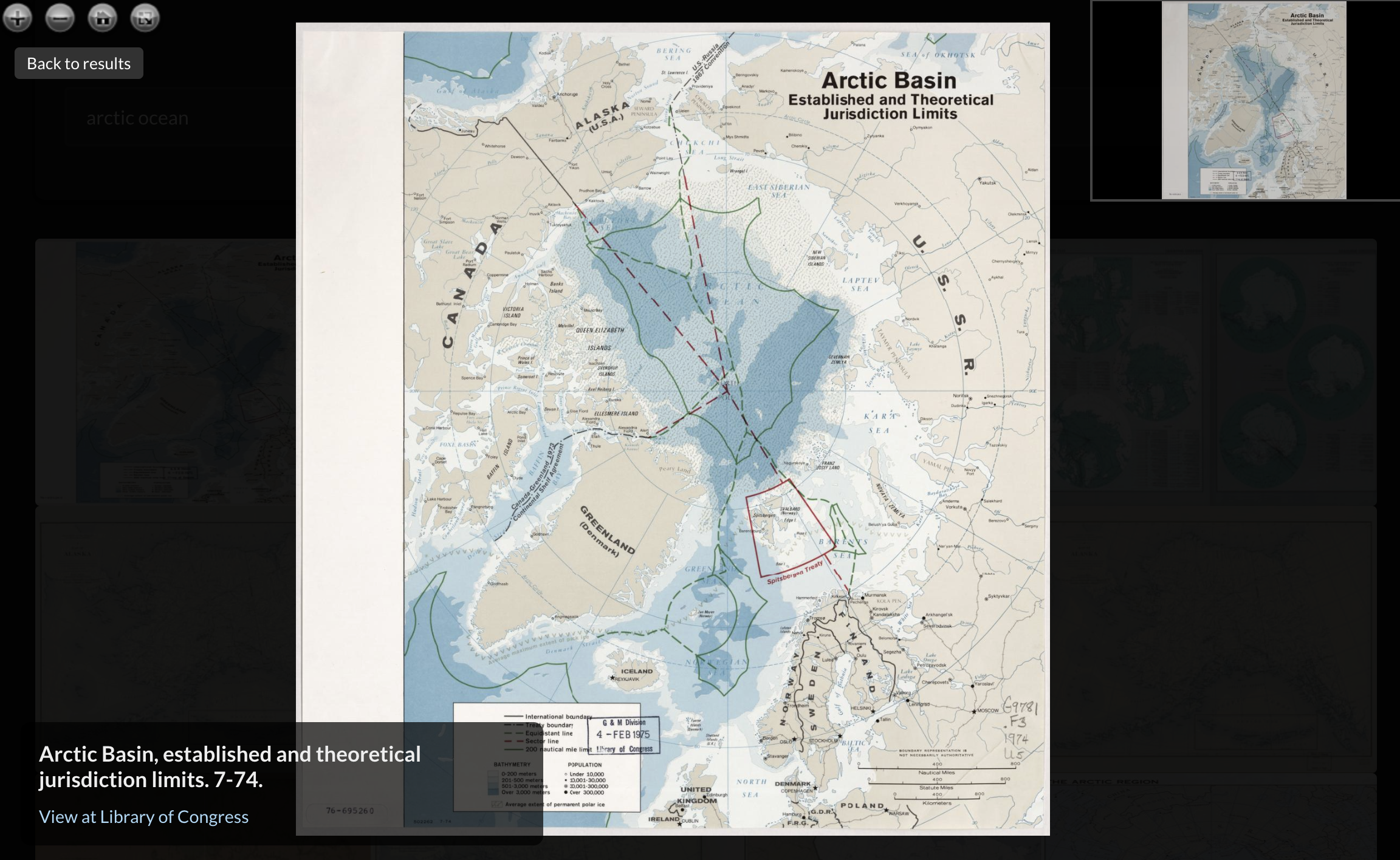}
    \caption{The lightbox view, built into the maps collection interface, enables detailed inspection of a historical map. This modal interface provides tools for zooming and panning, allowing for a detailed examination of a map's features.}
    \label{fig:interface_demo_03}
\end{figure}

The user-facing interface is built using React, providing an intuitive and responsive experience. As shown in Figure~\ref {fig:codebase_fe_structure_overview}, the architecture is centered around the \texttt{App.jsx} component, which serves as the primary container and state manager. This architecture enables several key features for the end-user, including:
\begin{itemize}
    \item \textbf{Search interaction}, supported by the \texttt{SearchBar.jsx} and \texttt{ImageUpload.jsx} components. These components provide interfaces for natural language queries and reverse image search, respectively. An example of the landing page, as shown in Figure~\ref{fig:interface_demo_01}, demonstrates both text and image search functionalities.
    \item \textbf{A gallery view for browsing collections}, which is rendered by the \texttt{SearchResult.jsx} component.
    This component renders a grid of thumbnails based on the search results, as shown in Figure~\ref{fig:interface_demo_02} for the query ``arctic ocean''.
    \item \textbf{Detailed image inspection}, provided by the \texttt{Lightbox.jsx} component. As demonstrated in Figure~\ref{fig:interface_demo_03}, this feature presents a high-resolution version of a selected item in a modal overlay.
\end{itemize}

For greater modularity and ease of reuse, the front-end is structured as independent React applications for each collection type (photographs, maps, and documents). Each application is self-contained, allowing developers to isolate and utilize a single front-end implementation for their specific needs. Additionally, this approach allows for collection-specific customization within a consistent structure; for example, while all collection types share common API call logic, the \texttt{frontend/documents/src/components} folder consists of a \texttt{PDFViewer.jsx} component tailored to its specific PDF content.

All communication with the back-end is handled by a dedicated service layer, \texttt{api.js}. When a user initiates a search query, the corresponding UI component notifies \texttt{App.jsx}, which then invokes the necessary function from the \texttt{api.js} service. This service manages the asynchronous API request and returns the data to \texttt{App.jsx}, which updates its state, triggering a re-render of the interface to display the results.

\subsubsection{Back-end}
The API server is implemented using FastAPI, a high-performance Python web framework. As depicted in Figure~\ref{fig:codebase_structure_overview}, the back-end logic within \texttt{src/backend} is divided into two core sub-directories: \texttt{api} and \texttt{services}. The \texttt{api} directory defines the public-facing endpoints that the front-end communicates with, while the \texttt{services} directory contains the core logic, such as the CLIP model inference and embedding management. This separation of concerns ensures maintainability.
The front-end interacts primarily with two main endpoints: \texttt{/api/search/text} for natural language queries and \texttt{/api/search/image} for reverse image search.
Both endpoints accept parameters for pagination, such as \texttt{limit} and \texttt{offset}, allowing for efficient loading of large result sets.
Upon receiving a request, the back-end processes the query and returns a ranked list of relevant items.
Each item in the response payload includes a unique identifier, its similarity score, and any associated metadata, providing the front-end with all necessary information for rendering.

To implement the embedding functionality, our system leverages the \textit{Transformers} library by \cite{huggingface_transformers}. 
This library provides a robust and efficient implementation of the CLIP model.
By building upon this widely adopted open-source tool, we ensure that our system is not only reliable but also easily extensible, allowing for future integration of other pre-trained models from the Hugging Face ecosystem.

The core of our back-end is the retrieval engine, which implements the procedure detailed by \cite{mahowald_lee}.
For any given query (either text or image), the system computes a CLIP embedding and retrieves the nearest neighbors from the pre-computed embeddings of the collection, ranked by cosine distance. 
Our principal contribution resides in the implementation of this methodology, transitioning it from its initial Jupyter Notebook prototype in \cite{mahowald_lee} to a production-ready system. This was accomplished through the design of a FastAPI application, wherein the retrieval process is delivered via a high-performance, non-blocking API endpoint. The FastAPI's asynchronous capabilities were crucial in this engineering effort, providing the necessary throughput to support scalable queries across hundreds of thousands of items with minimal latency.
\subsection{Working with the Digital Collections Explorer: A Tutorial}

Whether one is utilizing historical photographs, historic maps, or born-digital documents, the Digital Collections Explorer offers a streamlined setup process and scalability for customization with a wide range of digital collections. This section demonstrates how researchers can easily and efficiently set up the system to meet their specific collection requirements.

\subsubsection{System Setup}
  The system initialization process is designed to be straightforward by assigning the specific collection type as an argument. The following examples demonstrate the setup process for different collection types:

  \begin{enumerate}[label=\textbf{Type \arabic*.}, leftmargin=3cm, itemsep=1em]
    \item[\textbf{Photographs}]
    \begin{lstlisting}[basicstyle=\ttfamily\small]
    npm run setup -- --type=photographs
    \end{lstlisting}
    Configures a gallery interface with grid and masonry layouts, optimized for large-scale image browsing.

    \item[\textbf{Maps}]
    \begin{lstlisting}[basicstyle=\ttfamily\small]
    npm run setup -- --type=maps
    \end{lstlisting}
    Implements an OpenSeadragon viewer for high-resolution zoomable maps with smooth pan and zoom capabilities.

    \item[\textbf{Documents}]
    \begin{lstlisting}[basicstyle=\ttfamily\small]
    npm run setup -- --type=documents
    \end{lstlisting}
    Provides a temporal navigation interface with a document viewer optimized for PDFs extracted from web archives.
  \end{enumerate}
Each setup command automatically configures the appropriate front-end components and back-end services optimized for the specific collection type.

\subsubsection{Data Preparation and Embedding Generation}

For a given digital collection, we begin by placing the collection in the \texttt{data/raw/} directory. The system recursively retrieves images from subdirectories, so any existing directory structure is acceptable, so long as all of the images exist nested within \texttt{data/raw/}. The system supports common image formats, including JPG, PNG, TIFF, or PDFs (PDFs are split at the page level and converted to images as part of running this pre-processing pipeline). Embeddings are generated by running:
  \begin{lstlisting}[basicstyle=\ttfamily\small]
python -m src.models.clip.generate_embeddings
  \end{lstlisting}
This command processes all images in the \texttt{data/raw} directory and creates embeddings in the \texttt{data/embeddings} directory. We report embedding generation times for multiple collection examples in the next section of the paper on case studies, but we note that hundreds of thousands of images can be processed on a single MacBook Pro in hours.

\subsubsection{Starting the Server}
After embedding generation, the back-end server is launched to provide API endpoints for search and exploration by running the following command:
  \begin{lstlisting}[basicstyle=\ttfamily\small]
python -m src.backend.main
  \end{lstlisting}
The API server will then start at http://localhost:8000.

\subsubsection{Front-end Customization and Build Process}

To enable front-end customization, we have active development with hot reloading. Once the back-end server is up, starting the front-end development server can be accomplished with:
  \begin{lstlisting}[basicstyle=\ttfamily\small]
cd src/frontend/[photographs|maps|documents] 
npm run dev
  \end{lstlisting}
These commands start a front-end development server at http://localhost:5173 with hot-reloading enabled. The development server will automatically proxy API requests to the back-end at http://localhost:8000. For production deployment, front-end assets must be built using the following command:
  \begin{lstlisting}[basicstyle=\ttfamily\small]
npm run frontend-build
  \end{lstlisting}
Then restart the back-end server to serve the updated front-end assets. The build process is only required when deploying to production environments (such as cloud servers) or when generating optimized JavaScript bundles for enhanced performance. For local development and testing purposes, running the front-end development server is sufficient.

\subsubsection{Publicly Hosting a Digital Collections Explorer}
It is straightforward to host a web application of Digital Collections Explorer for public access on cloud services such as Amazon Web Services (AWS).
Although manual setup can be done by following the tutorial above, we strongly recommend this containerized approach due to its significant advantages in ensuring environmental consistency, simplifying dependency management, and enhancing security.
As a result, we provide a \texttt{Dockerfile} that serves as the cornerstone for both deployment and scientific reproducibility.
This Dockerfile encapsulates the application stack -- the Python back-end, the compiled JavaScript front-end, and all package dependencies. By doing so, it creates a portable image of the entire system. The general deployment process involves:
\begin{enumerate}
    \item \textbf{Build the Docker Image.} The image should be tailored to a specific collection type by passing the \texttt{--build-arg} flag during the build process:
\begin{lstlisting}[basicstyle=\ttfamily\footnotesize]
docker build --build-arg COLLECTION\_TYPE=<type> -t <image-tag> .
\end{lstlisting}
    \item \textbf{(Optional) Push to a Container Registry.} For distribution, the newly created image can be pushed to a container registry, such as Docker Hub or AWS ECR:
\begin{lstlisting}[basicstyle=\ttfamily\small]
docker push <image-tag>
\end{lstlisting}
This step is not required if the image is built directly on the target machine.
    \item \textbf{Run the Container.} Finally, the application is launched by running the container from the Docker image.
\begin{lstlisting}[basicstyle=\ttfamily\small]
docker run -p 8000:8000 -v ./data:/app/data <image-tag>
\end{lstlisting}
\end{enumerate}

\subsection{Public Demo}

For those who would like to experiment with an instantiation of Digital Collections Explorer, we host a public demo at: \url{https://www.digital-collections-explorer.com}. This demo supports searching over 562,842 map images from the Library of Congress -- one of our case study collections described in detail in the next section.

To create this live, interactive demonstration of this system, we deployed Digital Collections Explorer on an AWS EC2 instance using the following process. First, we built a Docker image on a MacBook Pro M4 and pushed it to our public Docker Hub repository: \url{https://hub.docker.com/repository/docker/hinxcode/digital-collections-explorer}. Following this, we provisioned an AWS EC2 c6gd.large instance ($\$0.08$/hour), pulled the image, and launched the application by running ``\texttt{docker run}''.

For this specific deployment, we diverged from the standard setup in two key ways. First, instead of using the embedding generation pipeline, we directly used the pre-computed embeddings provided by \cite{mahowald_lee}. Second, to augment the pre-computed embeddings with essential metadata, we developed a Library of Congress data preprocessing script. Our Python script, \texttt{create\_loc\_assets.py},\footnote{The script for Library of Congress maps preprocessing is included in the project repository at: \texttt{scripts/create\_loc\_assets.py}.} processes the original image identifiers and performs a record lookup against \texttt{merged\_files.csv} to generate two key assets: 1) a new index file, \texttt{item\_ids.pt}, which replaces the original identifiers with stable keys while preserving their sequence, and 2) a \texttt{metadata.json} file that maps each key to its corresponding metadata, including a direct link to the item's entry in the Library of Congress. These output files are then placed within the \texttt{/data/embeddings} directory, adhering to the file structure outlined previously.

\section{Discussion: Case Studies}

As detailed in Figure~\ref{fig:overview}, the Digital Collections Explorer employs a three-stage pipeline for collection exploration: 1) Data Preparation, 2) Embedding Generation, and 3) Search \& Exploration. In this section, we describe our case studies with photographs, maps, and born-digital documents using the Digital Collections Explorer.

\subsection{Data Preparation}
Collections are ingested into the system by placing images in a designated directory. For this study, we used four datasets to demonstrate the system's capabilities:
\begin{enumerate}
    \item A collection of 1,025 photographs provided by the photojournalist Christopher Morris.
    \item A large-scale collection of 129,386 photographs from the San Francisco Chronicle.
    \item The Library of Congress \href{https://www.loc.gov/item/2020445568/}{dataset} of 1,000 random .gov PDFs extracted from the Library of Congress web archives, amounting to 12,287 pages of PDFs in total (the value of searching these PDFs visually has been described by \cite{lee_owens_2021}).
    \item 562,842 images of maps held by the Library of Congress, retrieved using the Library of Congress API by \cite{mahowald_lee}.
\end{enumerate}

\subsection{Embedding Generation}

In Table \ref{tab:embedding_time}, we report the times to generate embeddings for the collections with a 2024 MacBook Pro M4 Chip with 10-Core CPU, 10-Core GPU, and 16GB; in Table \ref{tab:total_time}, we report the total processing times (including parsing, thumbnails generation, and embeddings generation) with the same machine. As reported, the Digital Collections Explorer can scale to hundreds of thousands of images in a tractable fashion. We note that these times do not scale precisely linearly for multiple reasons, including file size (and thus re-sizing during embedding generation) and different required pre-processing steps (such as PDF parsing).

\begin{table}[h]
\begin{center}
\begin{tabular}{|c|c|c|}
\hline
\textbf{Collection}                    & \textbf{Items} & \textbf{Embedding Time}
\\
\hline \hline
Library of Congress Maps & 562,842 map images & under 24 hours*  \\
\hline
San Francisco Chronicle Photo Collection   & 129,386 photographs   &  1 hour 27 minutes 32 seconds   \\
\hline
Library of Congress .gov PDF dataset & 1,000 PDFs (12,287 pages) & 8 minutes 57 seconds \\
\hline
Chris Morris Photo Collection & 1,025 photographs   & 4 minutes 5 seconds  \\
\hline
\end{tabular}
\caption{Embedding generation times with a 2024 MacBook Pro M4 Chip with 10-Core CPU, 10-Core GPU, and 16GB Unified Memory (*=reported by \cite{mahowald_lee} using similar hardware). We report failures on 75 images from the San Francisco Chronicle Photo Collection (0.058\%) and 17 Library of Congress PDFs (1.7\%).}\label{tab:embedding_time}
\end{center}
\end{table}

\begin{table}[h]
\begin{center}
\begin{tabular}{|c|c|c|}
\hline
\textbf{Collection}                    & \textbf{Items} & \textbf{Total Processing Time}
\\
\hline 
\hline 
San Francisco Chronicle Photo Collection   & 129,386 photographs   &  3 hours 20 minutes 19 seconds   \\
\hline
Library of Congress .gov PDF dataset & 1,000 PDFs (12,287 pages) & 28 minutes 24 seconds \\
\hline
Chris Morris Photo Collection & 1,025 photographs   & 10 minutes 41 seconds  \\
\hline
\end{tabular}
\caption{Total processing times (including parsing, thumbnails generation, and embeddings generation) with a 2024 MacBook Pro M4 Chip with 10-Core CPU, 10-Core GPU, and 16GB Unified Memory. Here, we omit the Library of Congress maps because we used the embeddings from \cite{mahowald_lee}.
}\label{tab:total_time}
\end{center}
\end{table}

\subsection{Search and Exploration}

\begin{figure*}[t!]
    \centering
    \begin{subfigure}[b]{0.9\textwidth}
        \centering
            \includegraphics[height=2.5in]{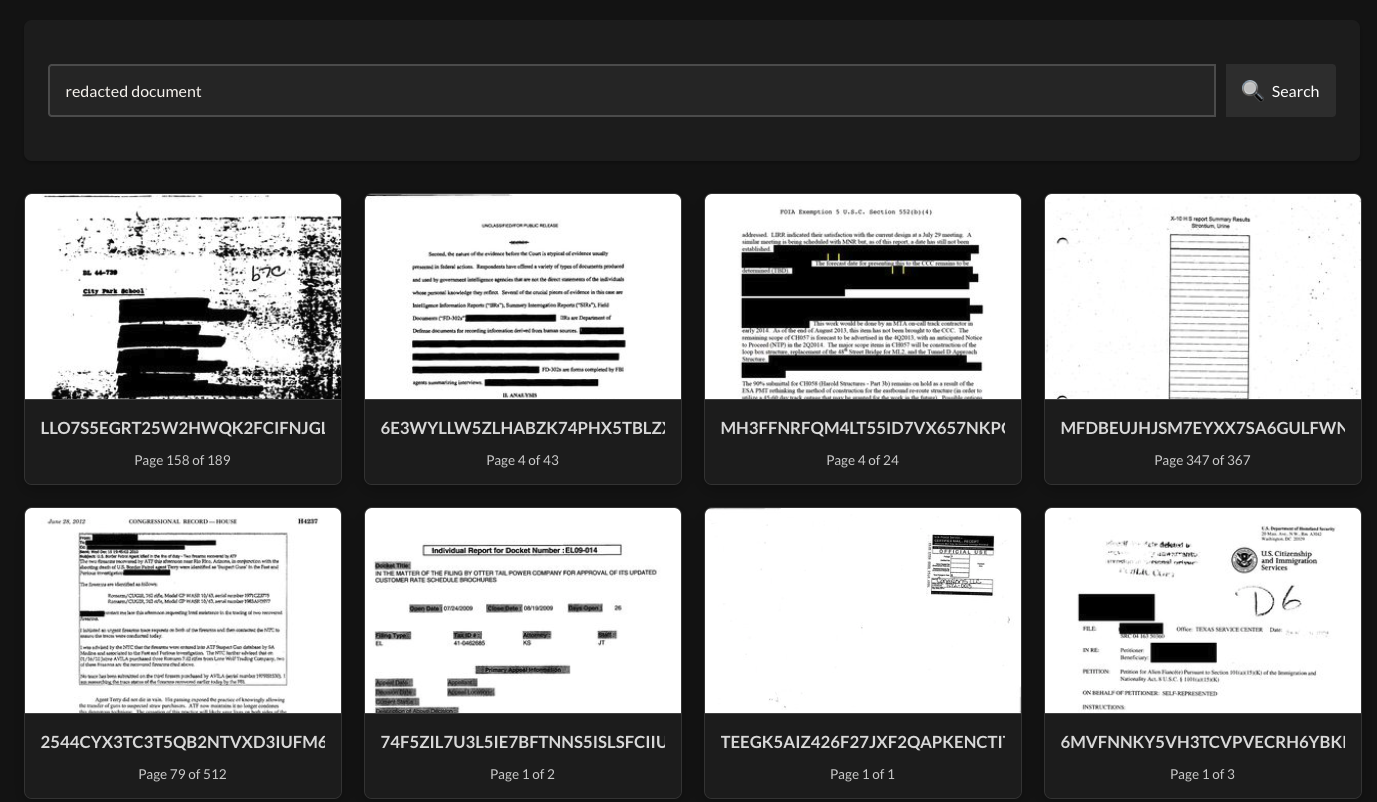}
        \caption{A search of ``redacted document''}\label{fig:redacted}
    \end{subfigure}
    ~ 
    \begin{subfigure}[b]{0.9\textwidth}
        \centering
            \includegraphics[height=2.5in]{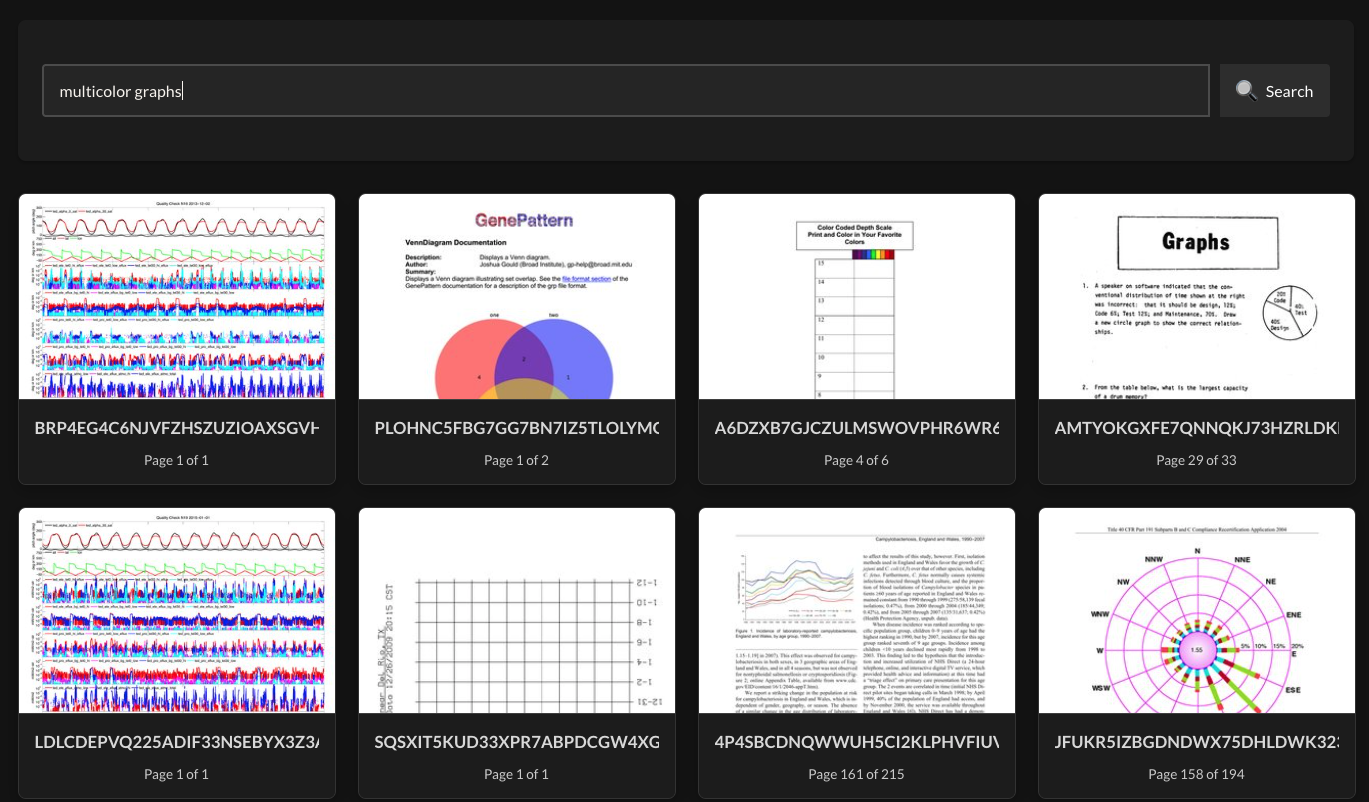}
        \caption{A search of ``multicolor graphs''}\label{fig:graphs}
    \end{subfigure}
    \caption{Search results for two different natural language queries across the 1,000 Library of Congress .gov PDFs demonstrating the effectiveness of semantic retrieval: (a) ``redacted documents,'' (Figure \ref{fig:redacted}) and (b) ``multicolor graphs'' (Figure \ref{fig:graphs}). The filenames shown refer to the PDF filenames (given by the hash in the Library of Congress web archives).}\label{fig:PDFs}
\end{figure*}

Here, we present example search results using two of our case studies: the Library of Congress maps and .gov PDFs, both of which are public domain collections (we have withheld screenshots of our other case studies due to copyright considerations).

In Figure \ref{fig:PDFs}, we present two natural language searches against the 1,000 .gov PDF dataset from the Library of Congress. Here, searches for ``redacted document'' and ``multicolor graphs'' result in ranking the PDF pages according to relevance to the search performed. As evidenced by these examples, we are able to query the visual features of the documents, rather than just their textual content.

\begin{figure*}[t!]
    \centering
    \begin{subfigure}[b]{0.9\textwidth}
        \centering
            \includegraphics[height=2.5in]{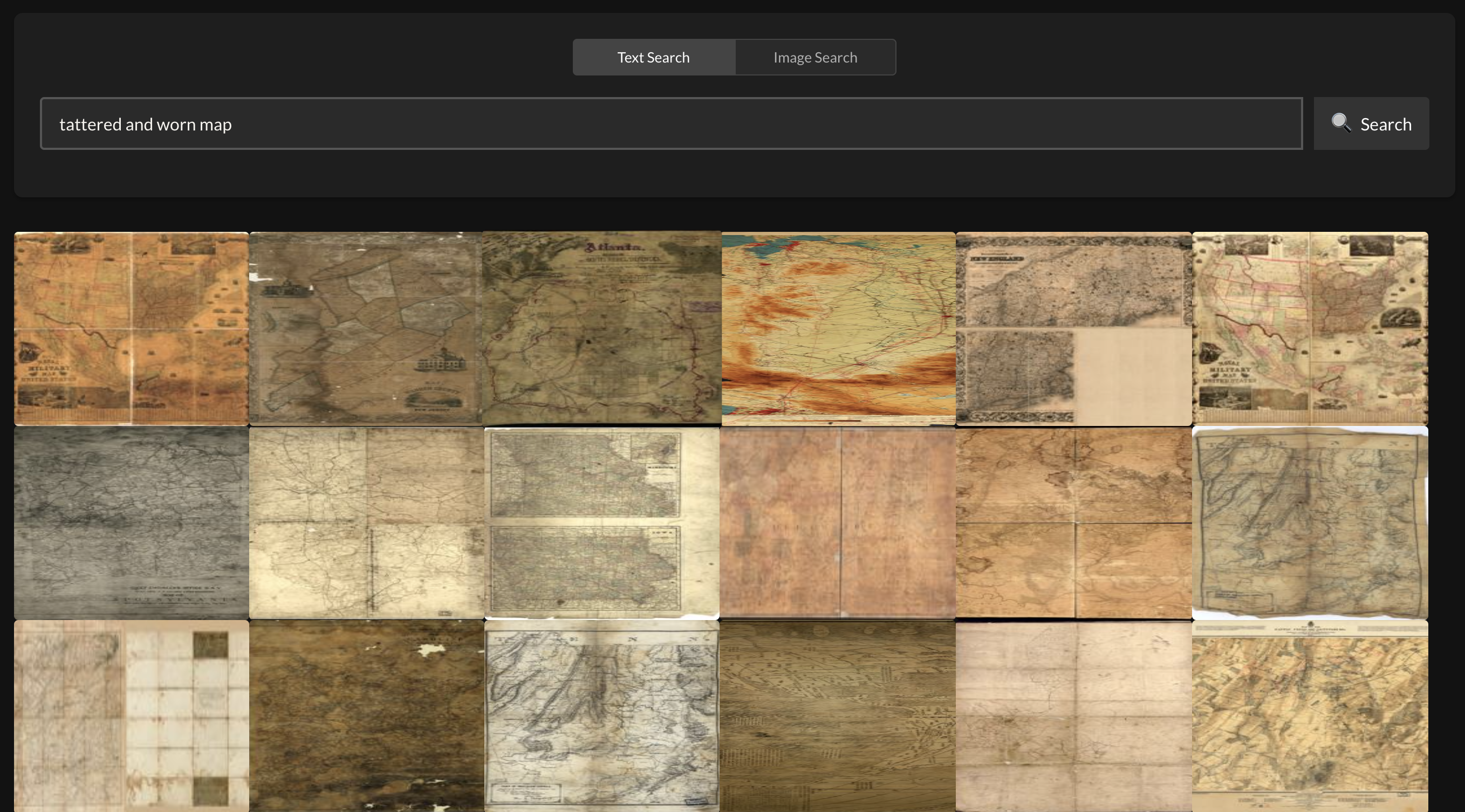}
        \caption{A search of ``tattered and worn map.''}\label{fig:tattered}
    \end{subfigure}
    ~ 
    \begin{subfigure}[b]{0.9\textwidth}
        \centering
            \includegraphics[height=3.12in]{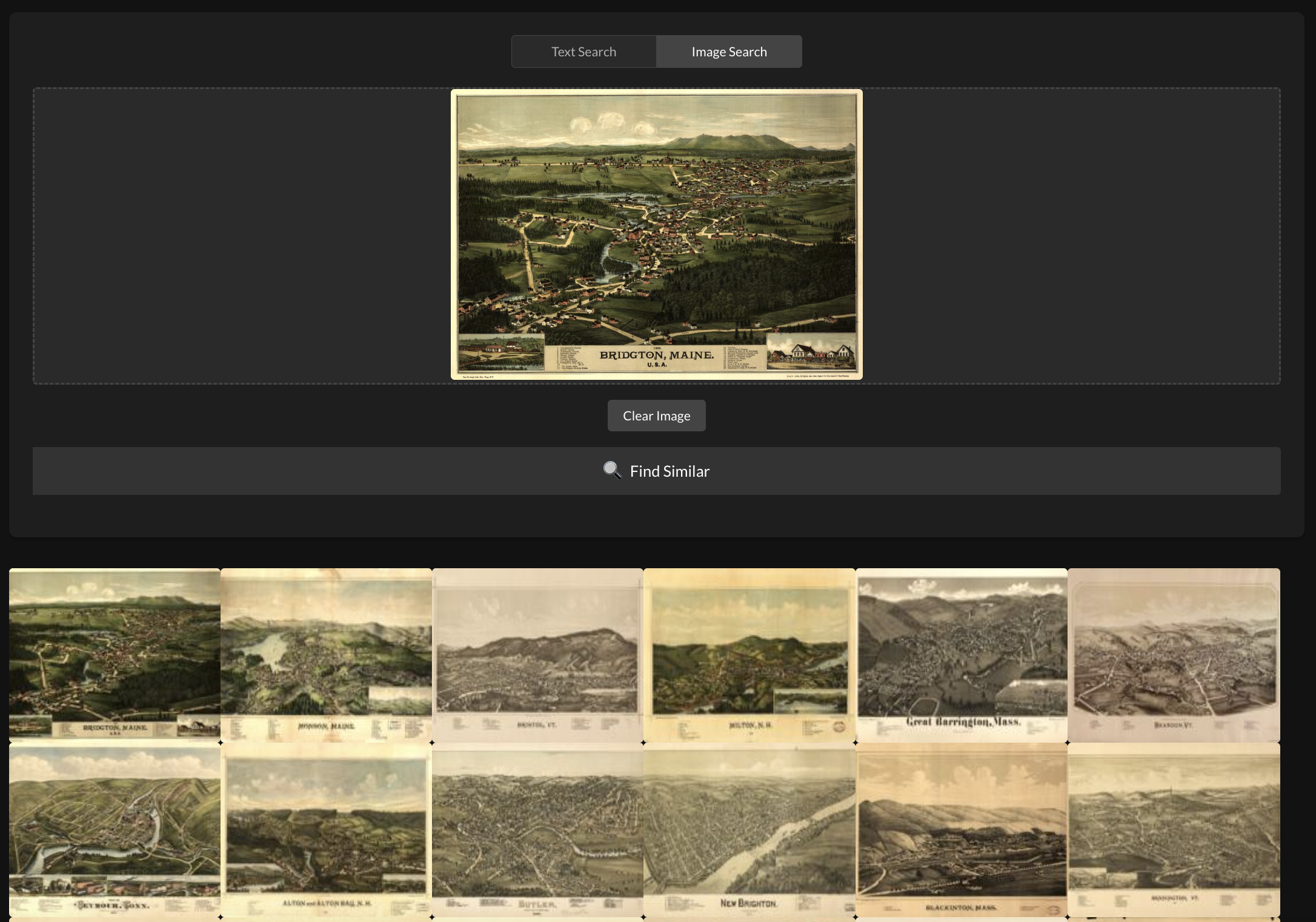}
        \caption{A reverse image search, with the input image shown at the top.}\label{fig:reverse}
    \end{subfigure}
    \caption{Searches against the 562,842 maps images from the Library of Congress API. Figure \ref{fig:tattered} shows a natural language search of ``tattered and worn map,'' and Figure \ref{fig:reverse} shows a reverse image search with a panoramic map of 1888 Bridgerton, Maine, from the Library of Congress's collections (\url{http://hdl.loc.gov/loc.gmd/g3734b.pm002434}). These results can be reproduced in our demo: \url{https://www.digital-collections-explorer.com}.}\label{fig:maps}
\end{figure*}

In Figure \ref{fig:maps}, we present searches against the 562,842 map images from the Library of Congress API. Figure \ref{fig:tattered} shows a natural language search of ``tattered and worn map''; we note that these results match the results from Figure 5a in \cite{mahowald_lee}, thereby confirming our ranking logic. This time, however, the searches can be performed in a production-ready user interface, rather than in Jupyter notebooks. Figure \ref{fig:reverse} shows a reverse image search returning relevant results. 
Any user can reproduce these searches in Figure \ref{fig:maps} and try others using our demo at: \url{https://www.digital-collections-explorer.com}.

We note that some searches work better than others. For example, the Digital Collections Explorer supports searching over visual content in the .gov PDFs - figures, images, etc. - but does not support semantically searching the text. For a more thorough investigation of the strengths and weaknesses in the search methods we employ, we refer the reader to \cite{mahowald_lee}.

\section{Conclusion \& Future Work}

In this paper, we have introduced our Digital Collections Explorer. With this open-source platform, researchers and practitioners can spin up a search interface on top of a digital collection of interest for enhanced visual discovery using both textual and visual inputs. Our work builds on the emerging body of research demonstrating the value of multimodal search and analysis for digital collections held by libraries, archives, and museums. Our Digital Collections Explorer extends this work by providing tooling to non-experts, enabling them to explore a digital collection in a multimodal fashion in just a few steps on a staff-issued laptop, such as current-generation MacBook Pro. Our platform is designed to scale to hundreds of thousands of images in this context. We have released the Digital Collections Explorer as open-source software under a CC-BY-4.0 license.

In order to preserve the privacy of digital collections, all steps of the Digital Collections Explorer can be run locally, from pre-processing to viewing, meaning that no data is transferred via proprietary APIs or publicly-visible endpoints. The Digital Collections Explorer is intended to be particularly useful as a method for exploring collections with little-to-no descriptive metadata.

Throughout this paper, we have introduced the system architecture, walked through a tutorial of how to use the Digital Collections Explorer, and presented case studies across maps, photojournalism collections, and born-digital PDFs extracted from web archives.  All of our code is available at: \url{https://doi.org/10.5281/zenodo.15744570}, and our publicly-available demo is available at: \url{https://www.digital-collections-explorer.com}.

While this paper demonstrates the core functionality of our Digital Collections Explorer, we hope to make a number of updates to the platform in the future in order to improve usability across a number of dimensions. First, we plan to provide support for additional input modalities, such as audio or video. Second, we plan to incorporate different multimodal models into our system beyond the one default CLIP model -- including models finetuned for cultural heritage collections. Third, we plan to experiment with models that are better-suited for searching text representations. Fourth, we hope to experiment with new modes of presenting metadata, as well as integrating external metadata sources and knowledge bases to enhance search capabilities. Fifth, we will plan to incorporate more options for GPU utilization in the embedding pipeline. Lastly, we will collect input and feedback from researchers and practitioners, which will inform future updates to the Digital Collections Explorer. We welcome contributions from the computational humanities and digital cultural heritage communities via submitting pull requests to our GitHub repository.

\paragraph{Acknowledgments}
 We are grateful to the University of Washington Information School for supporting this work. In particular, 
 we thank the Center for the Advancement of Libraries, Museums, and Archives (CALMA), for providing a research grant for the photojournalism portion of this work. We thank Chris Morris for his eagerness to share his photographs of Russia with us as an initial case study. We thank Kira Pollack for facilitating this collaboration, working with us, and providing invaluable feedback during this process. We also thank Nicole Fruge for providing us photographs from the San Francisco Chronicle as another case study for our system and for providing us invaluable feedback as well. Lastly, we thank Mark Lubell for facilitating conversations and collaborations surrounding the future of photojournalism archives.

\paragraph{Funding Statement}
This research was supported by the University of Washington Information School, including a grant from the Center for the Advancement of Libraries, Museums, and Archives.

\paragraph{Competing Interests} None.

\paragraph{Data Availability Statement} 
All code for the Digital Collections Explorer is available at: \url{https://doi.org/10.5281/zenodo.15744570}. In this GitHub repository, readers can find detailed instructions on how to install the Digital Collections Explorer and use it to view their own digital collections. The publicly-available datasets we have utilized in this paper are available as follows:
\begin{enumerate}
\item The 1,000 .gov PDF dataset by the Library of Congress is available at: \url{https://lccn.loc.gov/2020445568}.
\item The digitized maps are available through the Library of Congress's API.
\end{enumerate}

\paragraph{Ethical Standards}
The research meets all ethical guidelines, including adherence to the legal requirements of the study country. In addition, we have followed best practices from responsible AI in creating the Digital Collections Explorer.

\bibliographystyle{unsrt}  
\bibliography{bibliography}

\end{document}